\documentclass[aps,prb,showpacs,preprint,groupedaddress]{revtex4}
\usepackage[T1]{fontenc}
\usepackage[latin9]{inputenc}
\usepackage{amsmath}
\usepackage{graphicx}
\usepackage{setspace}
\usepackage{amssymb}

\begin{document}

\title{Solution of the Quantum Initial Value Problem
with Transparent Boundary Conditions}

\author{A. Puga and B.N.Miller\\
 \emph{Department of Physics and Astronomy, Texas Christian University,
Fort Worth, Texas 76129,USA}}
\begin{abstract}
Physicists have used billiards to understand and explore both
classical and quantum chaos. Recently, in 2001, a group at the
University of Texas introduced an experimental set up for modeling
the wedge billiard geometry called optical billiard in two
dimensions. It is worth mentioning that this experiment is more
closely related with classical rather than quantum chaos. The
motivation for the present work was born from the idea of laying the
foundations of a quantum treatment for optical billiards, named
{}``The Escape Problem'', by presenting the concept of a Transparent
Boundary Condition. We consider a {}``gas of particles'' initially
confined to a one dimensional box of length $L$, that are permitted
to escape. We find the solution of a Quantum Initial Value Problem
using a numerical method developed and entirely checked with an
exact, analytic method. The numerical method introduces a novel way
to solve a Diffusion Type Equation by implementing Discrete
Transparent Boundary Conditions recently developed by
mathematicians.
\end{abstract}
\maketitle

\section{INTRODUCTION}

The seminal model for understanding both classical and quantum chaos
is the billiard. Depending on different considerations, such as
shape and topology, a billiard model can present either stable or
chaotic behavior. The most celebrated versions of classical
billiards in two dimensions, in which chaos has been observed, are
the Sinai billiard\cite{Sinai}, Bunimovich stadium\cite{Bunimovich},
and Polygonal billiards\cite{Polygonal}. In all of these examples,
the shape of the boundaries plays a crucial role. In recent years,
physicists have detected chaotic behavior at the nano
scale\cite{Caos1} making an important impact in technology and, once
again, billiards have helped to model quantum devices\cite{Revsita}
such as nanotubes, quantum dots, etc. which are governed by quantum
theory.

A novel model of billiard is the one introduced in 1986 by Lehtihet
and Miller\cite{Miller} known as the \emph{{}``Wedge Billiard}.''
The model consists of a symmetrically inclined wedge of angle
$2\theta$ with respect to the direction of a constant gravitational
field $\mathbf{g}$ in which the particle is confined. They found
surprising properties when the parameter $\theta$ is changed. Later,
in 2001, Valery Milner\cite{Milner}, working in Mark Raizen's
laboratory at the University of Texas, introduced an experiment
referencing the Wedge Billiard model geometry and confirming the
properties mentioned above. This first experiment of billiards was
named {}``\emph{Optical Billiards}.'' It is worth mentioning that
this experiment is more closely related to \emph{classical} rather
than \emph{{}``quantum''} chaos\emph{.} While very low by normal
standards, the temperature they employed is not low enough to easily
exhibit any quantum effects. Consequently, a classical theoretical
model was adequate to obtain a good point of comparison.

A key element in the experiment is that, by removing a small segment
of the boundary, an escape route is provided at the billiard vertex.
As time progresses, the atoms, moving under the influence of
gravity, will eventually exit the billiard. A measure of the
influence of "chaotic" orbits is provided by the mean lifetime that
the atoms remain in the billiard. This scenario is mimicked in this
work by setting up the \emph{Escape Problem} (\emph{EP}). Currently
experimentalists are starting to probe the quantum regime.

The idea of the work presented here springs from laying the
foundation of a quantum treatment for \emph{optical billiards.}
However, the quantum mechanical problem is much more difficult since
the system is two-dimensional (equivalent to four dimensions in the
phase space), and not integrable. Thus analytical methods are not
available, and a viable numerical method for solving the Schrödinger
equation is required. Since it is extremely difficult to develop a
numerical method for a two dimensional integro-differential equation
with nonlocal transparent boundary conditions and skew boundaries,
this work will develop an essential numerical method for obtaining
the solution of the non-relativistic one-dimensional Schrödinger
equation.

The most demanding part of this work is the implementation of
\emph{Transparent Boundary Conditions}
(\emph{TBCs})\cite{Arnol1,Arnol2}. The \emph{TBCs}, recently
developed by mathematicians, arise in the necessity to deal
numerically with the natural infinite domain of the wave function.
That is, due to the limitation of the computer core size, the
infinite wave solution of the Schrödinger equation has to be solved
in a finite domain by imposing artificial boundary conditions. The
criterion of transparency is that the incident wave at the boundary
has the smallest reflection coefficient possible. If the solution
with these artificial conditions agrees with the infinite solution,
the artificial boundary conditions are said to be transparent. From
the several available approaches to derive \emph{TBCs}, this work is
based on the excellent treatment carried out by Anton
Arnold\cite{Arnol1}and his student Mathias
Earhardt\cite{Arnol2,tesis} that concerns the transport of a quantum
particle that enters one side of a finite domain and exits from the
opposite side.

In the following, in the next section, we first set up the
{}``\emph{Escape Problem}, i.e. the escape of a particle from a
finite region, as a \emph{Quantum Initial Value Problem}
(\emph{QIVP}). We then develop two different approaches for solving
the \emph{EP}. The first is the analytic method that provides the
certainty of the result. The second is the numerical method, where
an algorithm will be meticulously presented by introducing
\emph{Discrete Transparent Boundary Conditions (DTBC)\cite{tesis}},
based on the implicit \emph{Crank-Nicholson} method. Subsequently,
the consistency of both methods is confirmed in section III by
comparing the solutions for the real and imaginary parts of the wave
function at different times, showing an excellent agreement. In the
last section, conclusions are drawn from the work that has been
accomplished, stressing the valuable physical information provided
by the wave function as well as the numerical importance of the
method developed in this work.

\section{THEORY}

\subsection{DESCRIPTION OF THE SYSTEM.}

\begin{figure}[T]
\includegraphics[scale=1.0]{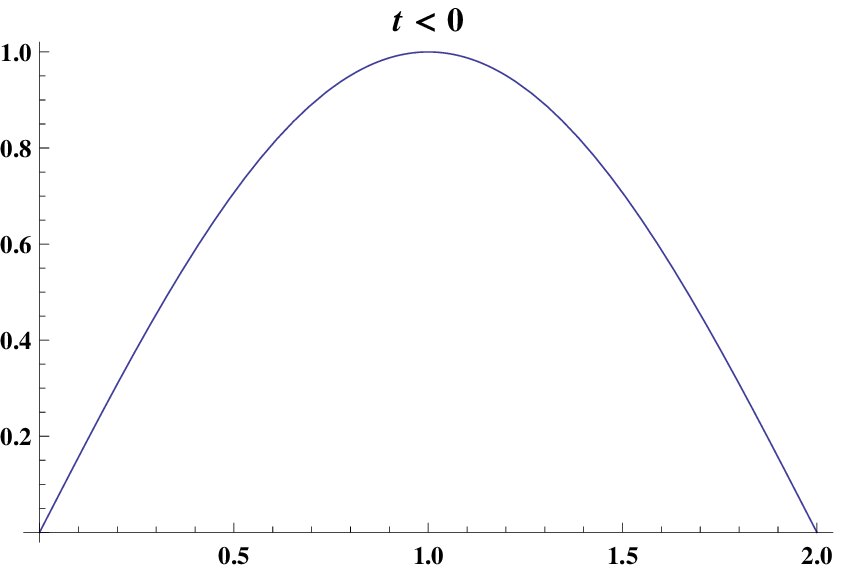}
\includegraphics[scale=1.0]{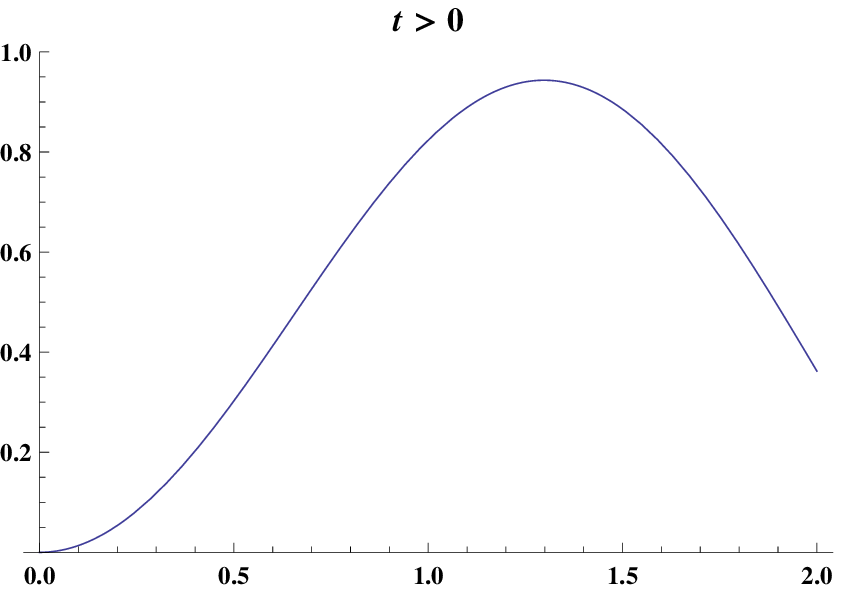}

\caption{\emph{\label{fig:Evolution EP}Top}: The \emph{EP} at time
$t<0$, the right boundary is closed. \emph{Bottom}: The \emph{EP} at
time $t>0,$ the right boundary is transparent allowing the wave
function to escape.}

\end{figure}

The system representing the one-dimensional \emph{Escape Problem}
(\emph{EP}) is a group of particles restricted inside a region of
size $L$ delimited by boundaries and free of external influences
where, abruptly, one of the boundaries becomes transparent, allowing
particles to escape. The group of particles is a dilute gas rarefied
enough to be considered as a {}``\emph{Knudsen
Gas}\textquotedblright{}\cite{Knudsen2} whose density is so small
that only the interactions with the boundaries are relevant. From
the perspective of quantum mechanics, the \emph{EP} is a
\emph{QIVP}. Thus, the intention is to solve the 1-D Schrödinger
equation from a given initial condition. The \emph{EP} is similar to
a diffusion type problem; it has the same structure as propagation
problems in which a partial differential equation is solved with the
aid of a known initial value. The evolution of the gas inside the
region $0<x<L$ will be determined by the wavefunction at the initial
time. To demonstrate the approach, here the initial wavefunction is
chosen to be the ground state of an \emph{Infinite Well Potential}
of length $L$. Therefore, the initial value or the initial condition
inside of the box is
\begin{equation}
\Psi_{I}\left(x,0\right)=\sin\left(kx\right)
\end{equation}
with $k=\frac{\pi}{L}$ . Of course, outside of the box the value of
the initial wave function vanishes. The evolution of the \emph{EP}
is represented in the figure \ref{fig:Evolution EP}.

\subsection{ANALYTICAL METHOD.}

The non-relativistic, time dependent, one-dimensional Schrödinger
equation is \begin{equation}
\frac{\partial\Psi\left(x,t\right)}{\partial t}=i\frac{\hbar}{2m}\frac{\partial^{2}\Psi\left(x,t\right)}{\partial x^{2}}\label{eq:SH eq}\end{equation}
 with the initial condition:\begin{equation}
\Psi\left(x,t\right)|_{t=0}=\Psi_{I}\left(x,0\right).\label{eq:Initial condition}\end{equation}

The \emph{Laplace Transform Method} is one of the many appropriate
techniques to solve differential equations. The first task is to
determine the \emph{Laplace transform} of the wave function and its
time derivative. For this particular example the definition of the
\emph{Laplace transform} is:

\begin{eqnarray*}
\phi\left(x,s\right) & = & \int_{0}^{\infty}\, e^{-st}\Psi\left(x,t\right)dt\end{eqnarray*}

or,

\begin{equation}
\phi\left(x,s\right)=\mathcal{L}\left\{ \Psi\left(x,t\right)\right\} .\label{eq:Laplace transform}\end{equation}
 By integrating by parts,

\begin{equation}
\mathcal{L}\left\{ \frac{d\Psi\left(x,t\right)}{dt}\right\} =s\phi\left(x,s\right)-\Psi_{I}\left(x,0\right).\label{eq:Laplace derivative transform}\end{equation}

After using equations \eqref{eq:Laplace transform} and \eqref{eq:Laplace derivative transform},
the Schrödinger equation \eqref{eq:SH eq}, after some algebraic re-arrangements,
becomes:\begin{equation}
\frac{\partial^{2}\phi\left(x,s\right)}{\partial x^{2}}+i\alpha s\phi\left(x,s\right)=i\alpha\Psi_{I}\left(x,0\right)\label{eq:Eq. Diferencial Lapace}\end{equation}
 with \begin{equation}
\alpha=\frac{2m}{\hbar}\label{eq:valor de alfa}\end{equation}

A \emph{Green's Function} can be used to solve the inhomogeneous
differential equations \eqref{eq:Eq. Diferencial Lapace}. The
following notation and definitions are necessary before using the
\emph{Green's Function}.

The Laplace transform of the \emph{Green's function} is denoted as:
\begin{equation}
\mathcal{L}\left\{ G\left(x,x',t\right)\right\} =g\left(x,x',s\right)\end{equation}
 where $G\left(x,x',t\right)$ is the solution in the non-transformed
space. The Green's function for equation \eqref{eq:Eq. Diferencial Lapace}
satisfies\begin{equation}
\frac{d^{2}g\left(x,x',s\right)}{dx^{2}}+i\alpha sg\left(x,x',s\right)=\delta\left(x-x'\right)\label{eq:Green's function}\end{equation}
 and its solution is given by\begin{equation}
\phi\left(x,s\right)=i\alpha\int_{0}^{\infty}dx'g\left(x,x',s\right)\Psi_{I}\left(x',0\right).\end{equation}
 The complete time dependent solution of the initial value problem
$\Psi\left(x,t\right)$ is obtained by inverting the Laplace transform:\begin{equation}
\Psi\left(x,t\right)=i\alpha\int_{0}^{\infty}dx'G\left(x,x',t\right)\Psi_{I}\left(x',0\right)\label{eq:funcion de onda usando Greeen}\end{equation}
 The remaining procedure consists in constructing the Green's function
$g\left(x,x',s\right)$ for our particular problem, establishing $G\left(x,x',t\right),$
and finally substituting it into equation \eqref{eq:funcion de onda usando Greeen}.

In order to construct the Green's Function $g\left(x,x',s\right)$ ,
beginning from equation \eqref{eq:Green's function} for $x\neq x'$
and assuming $ReP>0$ , the form of the postulated solution is
$g=e^{\pm Px}.$ Thus,

\begin{equation}
P=\sqrt{\frac{\alpha s}{2}}\left(1-i\right).\label{eq:valor de P}\end{equation}

For $x<x'$\begin{eqnarray}
g_{1} & = & Be^{Px}+Ce^{-Px}\nonumber \\
x=0 &  & g_{1}\left(0\right)=0\nonumber \\
& B+C=0 \\
g_{1} & = & B\left(e^{Px}-e^{-Px}\right)\label{eq:g uno}\end{eqnarray}

And for $x>x'$\begin{equation}
g_{2}=Ae^{-Px}\label{eq:g dos}\end{equation}

The following two conditions give $A$ and $B.$ For continuity $g_{1}=g_{2,}$
and for the discontinuity of the derivative $\frac{dg_{2}}{dx}|_{x'}-\frac{dg_{1}}{dx}|_{x'}=1.$

These two conditions lead to the value of:\begin{eqnarray}
B & =- & \frac{1}{2P}e^{-Px'}.\label{eq:constante B}\end{eqnarray}

Using this value of $B,$ the value for $A$ is derived as\begin{equation}
A=\frac{1}{2P}\left(e^{Px'}-e^{-Px'}\right).\end{equation}

The equation \eqref{eq:g uno} and the equation \eqref{eq:g dos}
can be rewritten to obtain:\begin{eqnarray}
for & x<x'\nonumber \\
g_{1} & = & -\frac{1}{2P}e^{-Px'}\left(e^{Px}-e^{-Px}\right)\\
for & x>x'\nonumber \\
g_{2} & = & -\frac{1}{2P}\left(e^{Px'}-e^{-Px'}\right)e^{-Px}.\end{eqnarray}

Alternatively,\begin{eqnarray}
g_{1} & = & g_{1a}+g_{1b}\nonumber \\
g_{1a} & = & -\frac{1}{2P}e^{-P\left(x'-x\right)}\\
g_{1b} & = & \frac{1}{2P}e^{-P\left(x+x'\right)}\end{eqnarray}

and,\begin{eqnarray}
g_{2} & = & g_{2a}+g_{2b}\nonumber \\
g_{2a} & = & -\frac{1}{2P}e^{-P\left(x-x'\right)}\\
g_{2b} & = & \frac{1}{2P}e^{-P\left(x+x'\right)}.\end{eqnarray}

In general, the inversion of the \emph{Laplace transform} of any of
the $g$ functions has to be done via contour integration in the
complex $s$ plane\emph{.} In addition, extensive mathematical tables
are also available with many worked out examples \cite{tablas}. This
section shows only one of the $g$ functions, $g_{1a}$ , as an
illustration of the use of these tables (the rest of functions
$g_{1b}$, $g_{2a}$, $g_{2b}$ are easy to obtain due to their
similarity). The following expressions can be found \cite{tablas}

\begin{eqnarray*}
f\left(s\right) &  & F\left(t\right)\\
\frac{1}{\sqrt{s}}e^{-k\sqrt{s}} &  & \frac{1}{\sqrt{\pi t}}e^{-\frac{k^{2}}{4t}}\end{eqnarray*}
 for $k>0$. Then, using \eqref{eq:valor de alfa} and \eqref{eq:valor de P}
it is easy to see that\begin{equation}
g_{1a}=\frac{e^{-\left(1-i\right)\sqrt{\frac{ms}{\hbar}}\left(x'-x\right)}}{2\left(1-i\right)\sqrt{\frac{ms}{\hbar}}}\end{equation}
 and its corresponding inverse \emph{Laplace transform} \begin{equation}
G_{1b}=\frac{e^{-\frac{1}{4t}\left[\frac{m}{h}\left(1-i\right)^{2}\left(x'+x\right)\right]}}{2\left(1-i\right)\sqrt{\pi t\frac{m}{\hbar}}}.\end{equation}

Similarly, it is easy to obtain

\begin{equation}
G_{2a}=-\frac{e^{-\frac{1}{4t}\left[\frac{m}{h}\left(1-i\right)^{2}\left(x'-x\right)\right]}}{2\left(1-i\right)\sqrt{\pi t\frac{m}{\hbar}}}\end{equation}

\begin{equation}
G_{2b}=\frac{e^{-\frac{1}{4t}\left[\frac{m}{h}\left(1-i\right)^{2}\left(x'+x\right)\right]}}{2\left(1-i\right)\sqrt{\pi t\frac{m}{\hbar}}}.\end{equation}

Defining the following constants $C\equiv\frac{1}{2\left(1-i\right)\sqrt{\frac{m\pi}{\hbar}}},$
and $a\equiv\frac{m}{2\hbar}$ the total \emph{Green's function} is
found to be:\begin{equation}
G\left(x.x',t\right)=\frac{C}{\sqrt{t}}\left[-e^{\frac{ia}{t}\left(x'-x\right)^{2}}+e^{\frac{ia}{t}\left(x'+x\right)^{2}}\right].\label{eq:resultante funcion de Green}\end{equation}

Once the Green's function is constructed by the above procedure, in
order to obtain the wave function it is necessary to substitute
equation \eqref{eq:resultante funcion de Green} into the equation
\eqref{eq:funcion de onda usando Greeen} resulting in the complete
solution:\begin{equation}
\Psi\left(x,t\right)=\label{eq:Fun Onda Final}\\
i\frac{\alpha
C}{\sqrt{t}}\int_{0}^{\infty}dx'\left(-e^{\frac{ia}{t}\left(x'-x\right)^{2}}+e^{\frac{ia}{t}\left(x'+x\right)^{2}}\right)\Psi_{I}\left(x',0\right)\nonumber
\end{equation}

Mathematica was employed for these computations and for plotting the
results.

\subsection{NUMERICAL METHOD.}

As mentioned earlier, the \emph{QIVP} can be treated as a
propagation problem. These are initial-value problems governed by a
parabolic Partial Differential Equation (\emph{PDE}) of first order
in the time. A familiar representation of a parabolic \emph{PDE} in
one dimension is the Diffusion Equation, \begin{equation}
f_{t}=\alpha\, f_{xx}.\label{eq:eq difusion}\end{equation}
 Subscripts $t$ and $xx$ represent the first time derivative and
second position derivative of the function $f$. The factor $\alpha$
in the equation is called the \emph{diffusivity} which is defined
by the system under investigation. The Schrödinger equation (in dimensionless
units) using the same notation as in equation \eqref{eq:eq difusion}
becomes\begin{equation}
i\Psi_{t}=-\frac{1}{2}\,\Psi_{xx}.\label{eq:Eq SH}\end{equation}

Except for the imaginary number $i$, the equation \eqref{eq:Eq SH}
is identical to \eqref{eq:eq difusion}. Therefore, it is
mathematically correct to proceed to solve \eqref{eq:Eq SH} with the
complex extension of the same tools used in the numerical method for
the Diffusion Equation. Numerical methods\cite{metodos} solve the
\emph{PDE} by transforming the integral problem into an algebraic
one that is computationally accessible. The \emph{Crank-Nicholson
method} is the preferred numerical algorithm used to solve the
Schrödinger equation as a diffusion type equation. It is an implicit
algorithm valid through second order in both space and time
coordinates, so it is very stable\cite{metodos}.

To make that equation equivalent to the escape problem, it is
necessary to employ the mathematical concepts of \emph{TBCs}. The
discretized domain of the solution is constructed using the
following finite difference grids
\begin{eqnarray}
x_{i} & = & \left(i-1\right)\triangle x\\
t^{n} & = & n\triangle t\;\left(\triangle t\, constant\right)\end{eqnarray}
 with $i=0,1,2,\ldots,imax$ and $n=1,2,\ldots,nmax$ . The domain
$D\left(x,t\right)$ is from $0$ to $L$ on the $x-axis$ and the
solution is marching in a positive time direction on the $y-axis.$
We introduce the following notation in the context of the finite
difference method grid point
$\left(i,t\right)\,\rightarrow\left(x_{i},t^{n}\right),$ function
$f\left(x_{i},t^{n}\right)\,\rightarrow f_{i}^{n},$ first time
derivative $\frac{\partial f_{i}^{n}}{\partial t}\,\rightarrow
f_{t}|_{i}^{n},$ and second space derivative
$\frac{\partial^{2}f_{i}^{n}}{\partial t^{2}}\,\rightarrow
f_{xx}|_{i}^{n}$.

The \emph{2nd Order Central Space} approximation of the second derivative
is:\begin{equation}
f_{xx}|_{i}^{n}=\frac{f_{i+1}^{n}-2f_{i}^{n}+f_{i-1}^{n}}{\triangle x^{2}}\label{eq:segunda derivada}\end{equation}

To obtain second order precision in the time, a key point of the
method is to estimate derivatives at half integral time steps. The
\emph{2nd Order Central Time} approximation of the first derivative
is given by:
\begin{equation}
f_{t}|_{i}^{n+\frac{1}{2}}=\frac{f_{i}^{n+1}-f_{i}^{n}}{\triangle
t}.\label{eq:primera dericada}
\end{equation}

The second space derivative is\begin{equation}
f_{xx}|_{i}^{n+\frac{1}{2}}=\frac{1}{2}\left(f_{xx}|_{i}^{n+1}+f_{xx}|_{i}^{n}\right).\label{eq:segunda
deriva en n+1/2}\end{equation}

The future time level occurs at $n,$ the past time level at $n-1$
and the algebraic relationship for the one-dimensional approximation
of the Schrödinger equation is:
\begin{equation}
i\left(\frac{\Psi_{i}^{n}-\Psi_{i}^{n-1}}{\triangle t}\right)=\\
-\frac{1}{4}\left(\frac{\Psi_{i+1}^{n}-2\Psi_{i}^{n}+\Psi_{i-1}^{n}}{\triangle
x^{2}}+\frac{\Psi_{i+1}^{n-1}-2\Psi_{i}^{n-1}+\Psi_{i-1}^{n-1}}{\triangle
x^{2}}\right).\nonumber
\end{equation}
 Then the \emph{Crank-Nicholson} difference equation is:\begin{equation}
-\Psi_{i+1}^{n}+\mathbb{C}\Psi_{i}^{n}-\Psi_{i-1}^{n}=\Psi_{i+1}^{n-1}-\mathbb{C}'\Psi_{i}^{n-1}+\Psi_{i-1}^{n-1}\label{eq:Iteracion Eq Difusion}\end{equation}
 with $\mathbb{C}=\left(2-i\rho\right)$ , $\mathbb{C}'=\left(2+i\rho\right)$
, and $\rho=\frac{4\triangle x^{2}}{\triangle t}.$

The algebraic relationship develops a tri-diagonal system represented
by the following equation

\begin{equation}
\widehat{A}\Psi^{n}=\vec{b},\label{eq:Sistema de ecuaciones para
SH}\end{equation} of a set of $imax$ simultaneous linear equations,
where $imax$ is the maximum number of grid points , $imax-1$
equations come from the interior scheme and one equation comes from
the \emph{Right Transparent Boundary Condition} (at $i=imax$ ). The
left boundary at $i=0$ is still a \emph{Dirichlet Boundary
Condition} and it is set to zero in order to force the function to
stay in the domain.

The code presented in this work was based on the design of \emph{DTBCs}
derived by the work of Arnold\cite{Arnol1,Arnol2} with the inclusion
of modifications that adapt \emph{DTBCs} to the particular example
of the escape problem. The original problem as conceived by Arnold
and derived in the dissertation of his student, Earhardt\cite{tesis},
evaluates the transport of a quantum particle that enters one side
of a finite domain and exits from the opposite side.

The \emph{Right TBC} is:

\begin{equation}
\Psi_{x}\left(L,t\right)=\label{eq:RTBC}\\
-\sqrt{\frac{2}{\hbar\pi}}e^{-i\frac{\pi}{4}}e^{-i\frac{V_{L}}{\hbar}t}\frac{d}{dt}\int_{0}^{t}\frac{\Psi\left(L,\tau\right)}{\sqrt{t-\tau}}e^{+i\frac{V_{L}}{\hbar}t}d\tau.\nonumber
\end{equation}

In contrast with \emph{Dirichlet conditions}, they are not stagnant
in time and the numerical method requires the storage of the past
history values at all time levels at the boundaries. The convolution
term on the right hand side of the equation \eqref{eq:RTBC} appears
as a fractional $\left(\frac{1}{2}\right)$ time derivative.
Derivation of the \emph{DTBCs} uses the \emph{Crank-Nicholson
scheme} at a discrete level. The discrete \emph{TBCs} for the one
dimensional Schrödinger equation for $n\geqslant1$ on the right is
at $j=J$
\begin{eqnarray}
 \Psi_{J-1}^{n}-l_{J}^{n}\Psi_{J}^{n}
& = &
\sum_{k=1}^{n-1}l_{J}^{\left(n-k\right)}\Psi_{J}^{k}-\Psi_{J-1}^{n-1}\label{eq:derecha}
\end{eqnarray}

with,

$l_{j}^{n}=\left(1-i\frac{R}{2}+\frac{\sigma_{j}}{2}\right)\delta_{n}^{0}+\left(1+i\frac{R}{2}+\frac{\sigma_{j}}{2}\right)\delta_{n}^{1}+\alpha_{j}e^{-in\varphi_{j}}\frac{P_{n}\left(\mu_{j}\right)-P_{n-2}\left(\mu_{j}\right)}{2n-1},$

$R=\frac{4}{\hbar}\frac{\left(\triangle x\right)^{2}}{\triangle t},$
$\varphi_{j}=\arctan\frac{2R\left(\sigma_{j}+2\right)}{R^{2}-4\sigma_{j}-\sigma_{j}^{2}},$

$\mu_{j}=\frac{R^{2}+4\sigma_{j}+\sigma_{j}^{2}}{\sqrt{\left(R^{2}+\sigma_{j}^{2}\right)\left(R^{2}+\left[\sigma_{j}+4\right]^{2}\right)}},$
$\sigma_{j}=2\left(\triangle x\right)^{2}V_{j},$

$\alpha_{j}=\frac{i}{2}\sqrt[4]{\left(R^{2}+\sigma_{j}^{2}\right)\left(R^{2}+\left[\sigma_{j}+4\right]^{2}\right)}e^{i\frac{\varphi_{j}}{2}}.$

The Right discrete boundary condition is derived at $i=imax-1$
\begin{eqnarray}
-\Psi_{imax}^{n}+\mathbb{C}\Psi_{imax-1}^{n}-\Psi_{imax-2}^{n} =\\
\Psi_{imax}^{n-1}-\mathbb{C}'\Psi_{imax-1}^{n-1}+\Psi_{imax-2}^{n-1}=b^{n-1}\left(imax-1\right)
\end{eqnarray}
where
\begin{equation}
i=L, \Psi_{imax}^{n}=\Psi\left(L,t\right),
\end{equation}
the \emph{Right Transparent Boundary Condition,} is part of the
system of equations. The \emph{Discrete Transparent Boundary
Condition} \eqref{eq:derecha} is:\begin{eqnarray}
\Psi_{imax-1}^{n+1}-l_{imax}^{\left(0\right)}\Psi_{imax}^{n+1} & =\nonumber \\
\sum_{k=1}^{n-1}l_{imax}^{\left(n-k\right)}\Psi_{imax}^{k}-\Psi_{imax-1}^{n}=b^{n-1}\left(imax\right).\label{eq:DTBC}
\end{eqnarray}
Note that $b^{n-1}\left(imax\right)$ will change
at every time level $n.$ This is the value at that boundary that
must be stored by the selected numerical method. For the purpose of
illustrating the procedure the first $n=3$ equations
are:\begin{eqnarray}
n=1\nonumber \\
\Psi_{imax-1}^{1}-l_{imax}^{\left(0\right)}\Psi_{imax}^{1} & =\nonumber \\
-\Psi_{imax-1}^{0}=b^{0}\left(imax\right)\\
n=2\nonumber \\
\Psi_{imax-1}^{2}-l_{imax}^{\left(0\right)}\Psi_{imax}^{2} & =\nonumber \\
l_{imax}^{1}\Psi_{imax}^{1}-\Psi_{imax-1}^{1}=b^{1}\left(imax\right)\\
n=3\nonumber \\
\Psi_{imax-1}^{3}-l_{imax}^{\left(0\right)}\Psi_{imax}^{3} & =\nonumber \\
l_{imax}^{2}\Psi_{imax}^{1}+l_{imax}^{1}\Psi_{imax}^{2}-\Psi_{imax-1}^{2}=b^{2}\left(imax\right)\end{eqnarray}

In matrix form, \begin{eqnarray}
\left[\begin{array}{cccccc}
\mathbb{C} & -1 & 0 & 0 & 0 & \cdots\\
-1 & \mathbb{C} & -1 & 0 & 0 & \cdots\\
0 & -1 & \mathbb{C} & -1 & 0 & \cdots\\
\vdots & \vdots & \ddots & \ddots & \ddots & \vdots\\
0 & 0 & \cdots & -1 & \mathbb{C} & -1\\
0 & 0 & 0 & \cdots & -1 & l_{imax}^{\left(0\right)}\end{array}\right]\left[\begin{array}{c}
\Psi_{2}^{n}\\
\Psi_{3}^{n}\\
\Psi_{4}^{n}\\
\vdots\\
\Psi_{imax-1}^{n}\\
\Psi_{imax}^{n}\end{array}\right]=\nonumber \\
\left[\begin{array}{c}
b\left(2\right)\\
b\left(3\right)\\
b\left(4\right)\\
\vdots\\
b\left(imax-1\right)\\
b\left(imax\right)\end{array}\right]\end{eqnarray}

A code called \emph{RTBC} was written in \emph{C} which employed the
\emph{LU Factorization} package to solve the complex valued
tri-diagonal system of simultaneous linear equations numerically.

\section{RESULTS.}

Since we use the implicit method called \emph{Crank-Nicholson} with
a degree of approximation through second order that is
unconditionally stable, there is no formal restriction on the step
sizes, and we expected little difference between the results
obtained by both methods. The real and imaginary parts of the wave
function are plotted for a set of different values of the time. For
simplicity, dimensionless units are used where $m=\hbar=1$ . In
these units we choose the length of the region to be $L=2.$ The set
begins with two small values of time and ends with a relatively
longer value compared with the value of the natural period. In
dimensionless units, $T=\frac{2\pi}{\omega}=5.092,$ for the first
ground state with a value of $\omega=1.233$.

\subsection{ANALYTICAL METHOD.}

The exact, time dependent wave function on the semi-infinite domain
is obtained by replacing the value of the initial condition in the
equation \eqref{eq:Fun Onda Final} as follows:
 \begin{equation}
 \Psi\left(x,t\right)=i\frac{\alpha
C}{\sqrt{t}}\int_{0}^{L}dx'\left(-e^{\frac{ia}{t}\left(x'-x\right)^{2}}+e^{\frac{ia}{t}\left(x'+x\right)^{2}}\right)\sin\left(kx'\right)
\end{equation}

After performing the integration using Mathematica Software, the explicit
form of the wave function is\[
\Psi\left(x,t\right)=\]
 \[
\frac{1}{\sqrt{2}}[\left(\left(\frac{1}{4}-\frac{i}{4}\right)\left(-1\right)^{\frac{1}{4}}e^{-\frac{i\pi\left(\pi t+2aLx\right)}{2aL^{2}}}\right)\]
 \[
(Erfi\left[\frac{\left(\frac{1}{2}+\frac{i}{2}\right)\left(-\pi t+aL\left(L-x\right)\right)}{\sqrt{a}L\sqrt{t}}\right]\]
 \[
-e^{\frac{2i\pi x}{L}}Erfi\left[\frac{\left(\frac{1}{2}+\frac{i}{2}\right)\left(\pi t+aL\left(L-x\right)\right)}{\sqrt{a}L\sqrt{t}}\right]\]
 \[
-e^{\frac{2i\pi x}{L}}Erfi\left[\frac{\left(\frac{1}{2}+\frac{i}{2}\right)\left(-\pi t+aL\left(L+x\right)\right)}{\sqrt{a}L\sqrt{t}}\right]\]
 \begin{equation}
+Erfi\left[\frac{\left(\frac{1}{2}+\frac{i}{2}\right)\left(\pi t+aL\left(L+x\right)\right)}{\sqrt{a}L\sqrt{t}}\right])]\end{equation}
 where $Erfi$ is the error function with imaginary argument \cite{tablas}.

\subsection{NUMERICAL METHOD.}

A code based on the \emph{Crank-Nicholson} algorithm was developed
to solve the \emph{TBC} problem. All of the code was written in the
C programming language and the calculations were performed on a
LINUX workstation using the GNU compiler. The major effort has gone
into adapting the techniques of \emph{TBCs} because it is
complicated by the fact that the boundary condition is non-local in
the time, but rather depends on a convolution related to fractional
calculus that cannot be treated by standard techniques. The standard
algorithm had to be modified to include the non-local, time
dependent boundary conditions, which now have to be updated at each
time step. To solve the Schrödinger equation on the finite interval,
the \emph{LU} decomposition method was employed. The function
libraries were obtained from the open source literature for the GNU
compiler.

\subsection{COMPARISON OF THE METHODS.}

\begin{figure}
\includegraphics[scale=1.0]{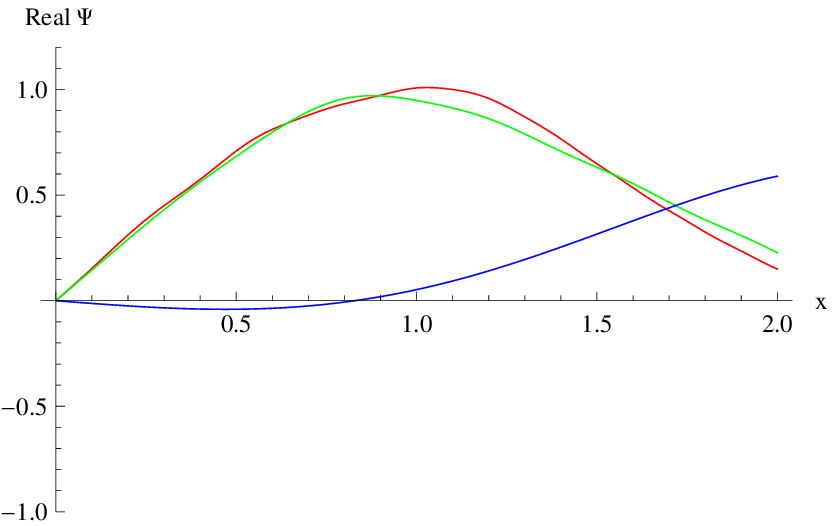}

\caption{\label{fig:Real analytically}Real part of the wave function obtained
analytically.}

\end{figure}

\begin{figure}
\includegraphics[scale=1.0]{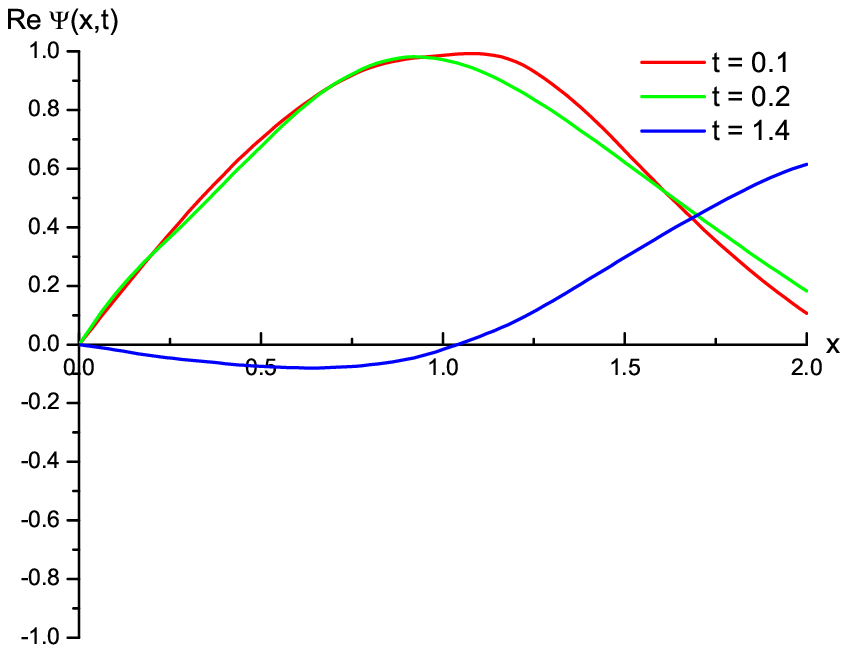}

\caption{\label{fig:Real numerically}Real part of the wave function obtained
numerically.}

\end{figure}

The first set of plots compare the real part of the wave function
for both methods at the same time steps $t=0.1$ , $t=0.2$ , and
$t=1.4$ as shown in figures \ref{fig:Real analytically} and \ref{fig:Real numerically}.

\begin{figure}
\includegraphics[scale=1.0]{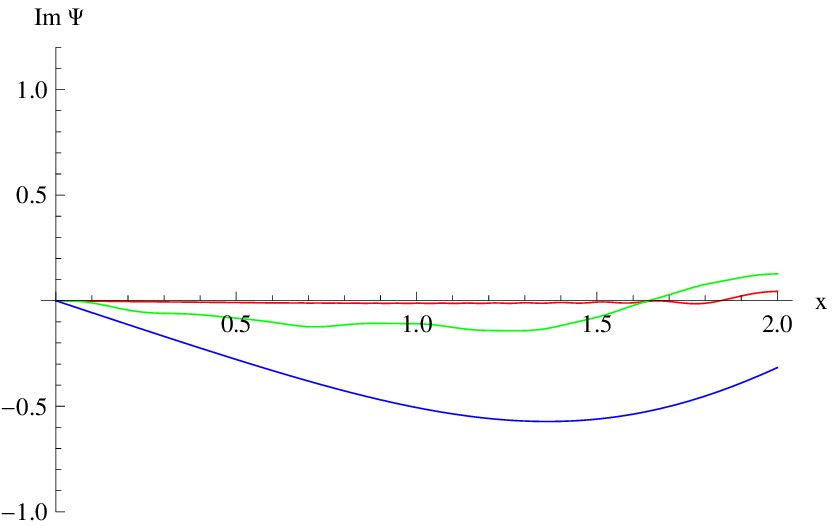}

\caption{\label{fig:Imaginary analytically}Imaginary part of the wave function
obtained analytically.}

\end{figure}

\begin{figure}
\includegraphics[scale=1.0]{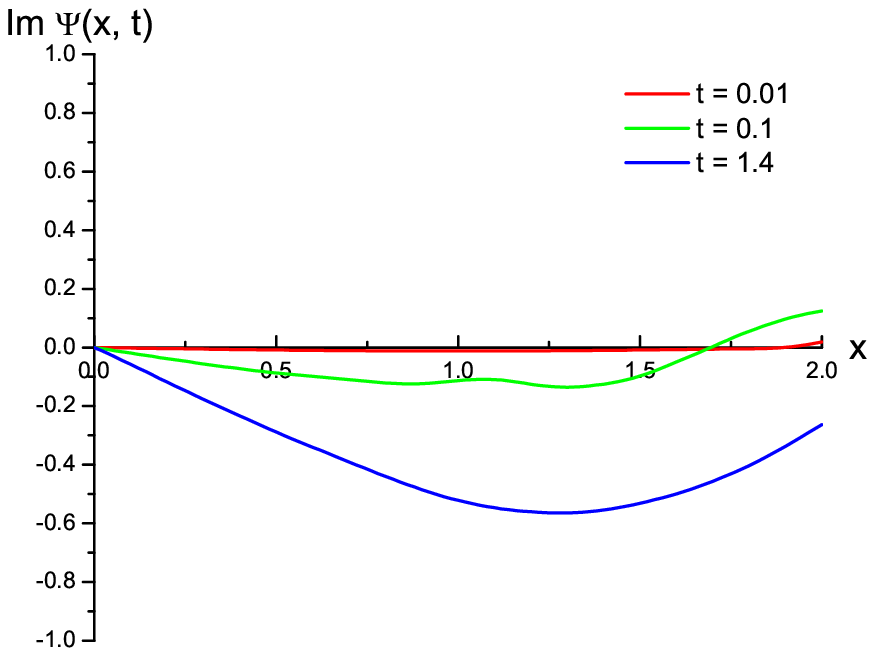}

\caption{\label{fig:Imaginary numerically}Imaginary part of the wave function
obtained numerically.}

\end{figure}

The second set of plots, figures \ref{fig:Imaginary analytically}
and \ref{fig:Imaginary numerically}, compare the imaginary part of
the wave function for both methods at the same time steps $t=0.01$
, $t=0.1$ , and $t=1.4.$

The comparison between the plots for each method shows an excellent
agreement as we expected. This confirms the adequate implementation
of the numerical method for a QIVP. Therefore, the basis can be established
for a more complex version of the \emph{RTBC}.

\section{CONCLUSIONS.}

In this work we investigated the problem of solving the time
dependent Schrodinger equation within a confined region when a
constraint is suddenly removed. We accomplished two important goals;
the first one is of physical significance. We determined the wave
function at any time $t,$ and therefore provided all the physical
information for the evolution of the \emph{EP} within the trap.
Therefore, we are able to extract important physical properties of
the system such as the mean fraction of mass inside the region,
which is called the survival probability, as well as the time
dependence of the properties of the particle kinetic energy,
potential energy, the total energy, etc.. Since it is assumed that
the particle density in the system is small, interactions can be
ignored (it is a \emph{Knudsen Gas}\cite{Knudsen2}), and multiplying
the survival probability by the total number of particles gives us a
quantitative value of mass inside the region.

The second conclusion concerns the numerical significance. We have
shown that the discretization developed by Arnold and Ehrhardt for
the solution of open boundary problems provides a viable numerical
approach for solving the time dependent Schrodinger equation
\cite{Arnol1,Arnol2,tesis} when a constraint is suddenly removed.
Although the present work considers a one-dimensional model that is
not sufficient to exhibit chaos, the code design forms the essential
basis of future research on non-integrable, two-dimensional billiard
models where chaos is present. The influence of chaos in these
models will be explored by measuring the survival probability of
particles in a trap with an open boundary.

The results obtained with the code presented here show an excellent
agreement with the analytic approach, offering reliability not only
in the result obtained but also in the novel numerical method used.
The success in the development of the numerical method opens up a
reasonable extension to higher dimensional models. Also, it offers
the possibility of understanding the quantum mechanical version of
billiard models that experience classical chaotic behavior. In spite
of the very low, by normal standards, temperature regime, a
classical theoretical model was still adequate for the analysis of
the Austin experiment. The quantum regime should also be accessible
experimentally at yet lower, but still attainable, temperatures.
Experimentalists are beginning to probe this regime.

\begin{acknowledgments}
The authors benefitted from the suggestions of M. Ehrhardt, A.
Arnold and K. Yawn, and the support of the division of Technology
Resources at Texas Christian University.
\end{acknowledgments}

\end{document}